\documentclass[twocolumn,prl,superscriptaddress,preprintnumbers,amsmath,amssymb]{revtex4}

\usepackage[paperwidth=210mm,paperheight=297mm,centering,hmargin={1.8cm,1.3cm}, vmargin={1.7cm,2.1cm}]{geometry}
\usepackage[pdftex]{graphicx}
\usepackage{dcolumn}
\usepackage{bm}
\usepackage{amssymb}
\usepackage{wasysym}
\usepackage{bibtopic}
\usepackage{sidecap}
\usepackage[rflt]{floatflt}
\usepackage{float}

\begin{document}

\def\Ef{$E_{\rm F}$}
\def\Eb{$E_{\rm B}$}
\def\Efmath{E_{\rm F}}
\def\Ed{$E_{\rm D}$}
\def\Tc{$T_{\rm C}$}
\def\kpara{{\bf k}$_\parallel$}
\def\kparamath{{\bf k}_\parallel}
\def\kperp{{\bf k}$_\perp$}
\def\Gbar{$\overline{\Gamma}$}
\def\Kbar{$\overline{K}$}
\def\Mbar{$\overline{M}$}
\def\BiTe{Bi$_2$Te$_3$}
\def\BiSe{Bi$_2$Se$_3$}
\def\SbTe{Sb$_2$Te$_3$}
\def\Ed{$E_{\rm D}$}
\def\invA{\AA$^{-1}$}

\title {Ultrafast spin polarization control of Dirac fermions in topological  insulators}
\author{J. S\'anchez-Barriga}
\affiliation{Helmholtz-Zentrum Berlin f\"ur Materialien und Energie, Albert-Einstein-Str. 15, 12489 Berlin, Germany}
\author{E. Golias}
\affiliation{Helmholtz-Zentrum Berlin f\"ur Materialien und Energie, Albert-Einstein-Str. 15, 12489 Berlin, Germany}
\author{A. Varykhalov}
\affiliation{Helmholtz-Zentrum Berlin f\"ur Materialien und Energie, Albert-Einstein-Str. 15, 12489 Berlin, Germany}
\author{J. Braun}
\affiliation{Department Chemie, Ludwig-Maximilians-Universit\"at M\"unchen, Butenandtstr. 5-13, 81377 M\"unchen, Germany}
\author{L. V. Yashina}
\affiliation{Department of Chemistry, Moscow State University, Leninskie Gory 1/3, 119991, Moscow, Russia}
\author{R. Schumann}
\affiliation{Max-Born-Institut, Max-Born-Str. 2A, 12489 Berlin, Germany}
\author{J. Min\'ar}
\affiliation{Department Chemie, Ludwig-Maximilians-Universit\"at M\"unchen, Butenandtstr. 5-13, 81377 M\"unchen, Germany}
\affiliation{New Technologies Research Centre, University of West Bohemia, Univerzitni 2732, 306 14 Pilsen, Czech Republic}
\author{H. Ebert}
\affiliation{Department Chemie, Ludwig-Maximilians-Universit\"at M\"unchen, Butenandtstr. 5-13, 81377 M\"unchen, Germany}
\author{O. Kornilov}
\affiliation{Max-Born-Institut, Max-Born-Str. 2A, 12489 Berlin, Germany}
\author{O. Rader}
\affiliation{Helmholtz-Zentrum Berlin f\"ur Materialien und Energie, Albert-Einstein-Str. 15, 12489 Berlin, Germany}

\begin{abstract}

{\bf Three-dimensional topological insulators (TIs) are characterized by spin-polarized Dirac-cone surface states that are protected from backscattering by time-reversal symmetry. Control of the spin polarization of topological surface states (TSSs) using femtosecond light pulses opens novel perspectives for the generation and manipulation of dissipationless surface spin currents on ultrafast timescales. Using time-, spin-, and angle-resolved spectroscopy, we directly monitor for the first time the ultrafast response of the spin polarization of photoexcited TSSs to circularly-polarized femtosecond pulses of infrared light. We achieve all-optical switching of the transient out-of-plane spin polarization, which relaxes in about 1.2 ps. Our observations establish the feasibility of ultrafast optical control of spin-polarized Dirac fermions in TIs and pave the way for novel optospintronic applications at ultimate speeds.}
\end{abstract} 

\maketitle

The emerging field of ultrafast spintronics in condensed-matter physics relies on the possibility of achieving efficient control of pure spin-currents, spin-polarized electrical currents and spin-configurations on ultrafast timescales \cite{Wolf-Science-2001, Das-Sarma-RMP-2004, Carva-NatPhys-2014}. One alternative for this purpose is the use of technologically relevant information-storage devices which are composed by ferromagnetic layers, and utilize laser-assisted switching for ultrafast remagnetization. For instance, the use of circularly-polarized femtosecond (fs) laser pulses has been established as a promising route to excite and coherently control spin dynamics in magnets without involving spin precession or external magnetic fields \cite{Rasing-RMP-2010}. The efficiency of such magnetic devices might be even enhanced by manipulation of isolated spins with pulses as short as the timescale of the exchange interaction, and within a single-photon shot \cite{Rasing-RMP-2010}. 

On the other hand, the growing demand of low-power consumption by using the least possible current in such devices, has motivated in recent years a completely different avenue to achieve generation and control of pure spin currents at ultimate speeds. This alternate pathway relies on the possibility of controlling carrier spins in low-dimensional systems through the spin-orbit interaction \cite{Rojas-NatComm-2013, Manchon-NatPhys-2014, Pesin-NatMat-2012, Winkler-Springer-2003}. The process of spin-current generation and its manipulation on sub-picosecond (ps) timescales is based on non-magnetic materials, and control of the electron spin is achieved solely by optical means \cite{Das-Sarma-RMP-2004, Optical-Orientation-1984}. This method also allows for the generation of spin-polarized electrical currents where in addition to the spin there is net flow of charge. With the advent of new classes of electronic materials such as topological insulators (TIs) \cite{Hasan-RMP-2010, Moore-Nature-2010, Fu-PRL-2007}, this unique route towards ultrafast optical control of spin currents and spin-polarized electrical currents appears to be very promising as it might lead to much lower energy consumption as compared to devices entirely composed by ferromagnetic layers. The key difference resides on the fact that, while insulating in the bulk due to strong spin-orbit coupling, the surface electronic structure of TIs is characterized by spin-polarized Dirac-cone topological surface states (TSSs) that are protected by time-reversal symmetry, hence by unprecedented properties such as forbidden backscattering \cite{Hsieh-Nature-09, Hsieh-Science-09, Roushan-Nature-2009}. This phenomenon opens the way for the realization of dissipationless spin-current devices or ultrafast information transport based on spin-polarized electrical currents on the surface of these materials \cite{Pesin-NatMat-2012}. For instance, it has been recently shown that such spin-polarized electrical currents flowing across TI surfaces can be utilized to exert spin-transfer torque effects in adjacent ferromagnetic layers, being the size of the torque greater than the one induced by any other material so far \cite{Mellnik-Nature-2014}.

In this context, generation of ultrafast spin-polarization transport in TIs means the creation of a nonequilibrium spin population of charge carriers above the Fermi level in sub-ps timescales \cite{Gedik-2011-PRL-Kerr, Gedik-2012-NatNanotech-photocurrents}. The unique response of photoexcited TSSs to fs light pulses has been demonstrated in a number of different experiments \cite{Gedik-2011-PRL-Kerr, Gedik-2012-NatNanotech-photocurrents, Sobota-PRL-2012-Bulk-Reservoir, Gedik-2012-PRL-phonons, Crepaldi-2012-PRB-phonons, Gedik-2013-Science-Floquet, Sobota-PRL-2013, Reimann-2014-PRB-phonons, Sobota-PRL-2014-Oscillations, Hajlaoui-NatComm-2014-e-h-asym, Sobota-JESRP-2014, Luo-NanoLett-2013-increased-e-ph}, primarily using time- and angle-resolved photoemission spectroscopy (tr-ARPES) \cite{Sobota-PRL-2012-Bulk-Reservoir, Gedik-2012-PRL-phonons, Crepaldi-2012-PRB-phonons, Gedik-2013-Science-Floquet, Sobota-PRL-2013, Reimann-2014-PRB-phonons, Sobota-PRL-2014-Oscillations, Hajlaoui-NatComm-2014-e-h-asym, Sobota-JESRP-2014}. Based on a pump-probe scheme providing access to transient population and depopulation dynamics, tr-ARPES has become the most powerful tool in systematically revealing the nonequilibrium dynamics of TI materials in energy and momentum space. Recent studies utilizing this technique have primarily identified bulk-assisted electron-phonon scattering as the underlying mechanism responsible for the ultrafast momentum relaxation of photoexcited carriers in TIs  \cite{Sobota-PRL-2012-Bulk-Reservoir, Gedik-2012-PRL-phonons}. It has been shown that following optical excitation by intense infrared fs pulses, this mechanism leads to an ultrafast response of both bulk and surface electronic structure that is characterized by coherent-phonon oscillations at terahertz frequencies \cite{Sobota-PRL-2014-Oscillations}. In addition, exotic topological quantum phases such as transient Floquet-Bloch states have been experimentally observed, and coherent control of these photon-dressed surface bands has been achieved by circularly-polarized ultrashort midinfrared pulses with energies below the bulk band gap \cite{Gedik-2013-Science-Floquet}.

Although extensive spin-resolved ARPES studies on various TI compounds have successfully identified the helical spin texture of TSSs \cite{Hsieh-Nature-09, Hsieh-Science-09, Souma-PRL-2011-spin, Pan-PRL-2011-spin, Jozwiak-PRB-2011-spin, Pan-PRB-2013-spin}, or the conditions for its manipulation using circularly-polarized photons in equilibrium \cite{Louie-PRL-2012-Theo-Spin-Flip, Jozwiak-NP-2013-Spin-flip, Sanchez-Barriga-PRX-2014}, the critical response of the spin properties of photoexcited TSSs to the incident light and its polarizations on ultrafast timescales has remained unexplored, mainly due to the lack of spin resolution in previous time-resolved experiments. Based on indirect methods such as time-resolved reflectivity or Kerr rotation, it has been suggested that relaxation of spin and charge degrees of freedom of photoexcited TSSs occurs on very different timescales \cite{Gedik-2011-PRL-Kerr}. The experiments indicate that spin relaxes much faster than charge, in particular within the first few fs following optical excitation, which corresponds to the timescale of electron-electron scattering \cite{Sobota-PRL-2012-Bulk-Reservoir, Gedik-2012-PRL-phonons, Crepaldi-2012-PRB-phonons, Othonos-JAP-1998}. Such an effect enabled the observation of light-induced electrical currents flowing across TI surfaces by means of transport measurements, and without the need of time resolution \cite{Gedik-2012-NatNanotech-photocurrents}. Specifically, reversal of these currents with the helicity of circularly-polarized fs pulses allowed to attribute them to spin-polarized electrical currents originating from the unique helical spin texture of TSSs \cite{Gedik-2012-NatNanotech-photocurrents}. However, a direct microscopic observation of the ultrafast spin-polarization dynamics of TSSs shedding light on the time evolution and underlying origin of these helicity-dependent photocurrents, is still missing.
 
Here, for the first time, we provide the key experimental evidence that allows us to directly determine the ultrafast nonequilibrium dynamics of the momentum-space spin configurations induced on TI surfaces using circularly-polarized light. We perform spin-resolved tr-ARPES measurements on the naturally p-doped TI \SbTe, which is characterized by completely unoccupied TSS bands in equilibrium. We demonstrate all-optical control of the nonequilibrium spin-polarization of Dirac fermions on ultrafast timescales and further identify the underlying mechanism that triggers the spin-relaxation process following optical excitation.\\

\noindent{\bf Nonequilibrium bulk and surface dynamics}

\noindent{To} study the temporal evolution of the transient populations in the unoccupied band structure of \SbTe, we perform pump-probe experiments under the two-color optical excitation scheme shown in Fig. 1a. 
Infrared and ultraviolet fs-laser pulses of 1.5 and 6 eV photon energy play the role of pump and probe beams, respectively, both incident on the sample at an angle of $\phi=45^{\circ}$ with respect to the surface normal. The time resolution of the present experiment is $\sim$200 fs, and the pump fluence $\sim$300 $\mu$J/cm$^{2}$. We optically excite electrons from the occupied states into unoccupied bands near the Fermi level using circularly-polarized pump pulses of positive or negative helicity (C+ or C-). Subsequently, we monitor the ultrafast response of the transiently modified electron distribution by emitting photoelectrons above the vacuum barrier using linearly s-polarized probe pulses at variable time delays $\Delta$t.

Figure 1b shows low-energy tr-ARPES measurements recorded along the \Gbar\Kbar\ direction of the \SbTe\ surface Brillouin zone (SBZ). The spectra are taken with this direction oriented perpendicular to the light incidence yz plane in Fig. 1a, and with infrared-pump pulses of C+ polarization. At negative delays ($\Delta$t=-1ps), we probe the electronic structure in equilibrium, as evidenced by the lack of photoemission intensity above the Fermi level. Following optical excitation, two transient electron populations in the initially unoccupied TSS and bulk-conduction band (BCB) are clearly observed. These populations rapidly thermalize to a nonequilibrium Fermi-Dirac distribution by means of electron-electron scattering processes \cite{Sobota-PRL-2012-Bulk-Reservoir, Gedik-2012-PRL-phonons, Crepaldi-2012-PRB-phonons}. Subsequently, the electronic system undergoes an ultrafast relaxation and completely disappears through an avalanche of hot electrons within $\sim$3 ps (right panel in Fig. 1b). This process is characterized by a significant energy transfer to the phonon system, which occurs via electron-phonon scattering \cite{Sobota-PRL-2012-Bulk-Reservoir, Gedik-2012-PRL-phonons, Sobota-JESRP-2014, Othonos-JAP-1998}. 

\begin{figure*}
\centering
\includegraphics [width=0.8\textwidth]{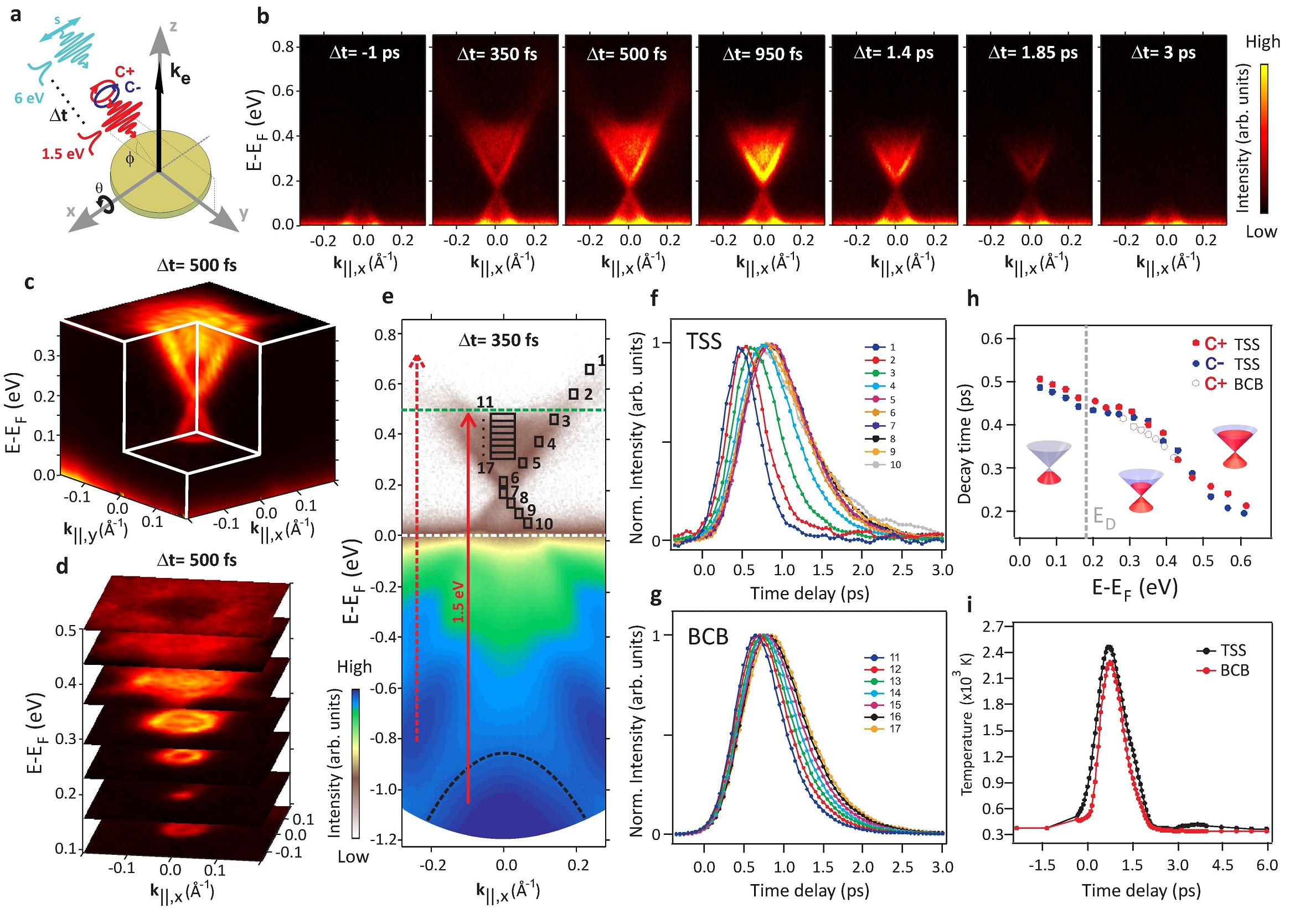}
\caption{{\bf Ultrafast dynamics of Dirac fermions following optical excitation by circularly-polarized infrared pulses.} (a) Experimental geometry. (b) tr-ARPES spectra obtained along the \Gbar\Kbar\ direction of the SBZ using pump pulses of positive helicity (C+). (c, d) Full-ARPES mapping 500 fs after photoexcitation. The Dirac cone exhibits circular constant-energy contours that evolve into warped regions with increasing energy. (e) Observation of direct optical transitions from deeper-lying bulk states into Dirac bands above the Fermi level. The TSS bands right below and in the immediate vicinity of the Fermi level strongly overlap with bulk-valence band states. (f, g) Normalized tr-ARPES intensities of (f) the TSS and (g) BCB as a function of pump-probe delay. The curves are extracted from the small energy-momentum windows shown in (e). (h) Energy-dependence of the TSS and BCB decay times for opposite pump-pulse helicities. (i) Transient electronic temperatures.}
\label{Dynamics}
\end{figure*}

Such a rapid electron recombination evolves mainly via two cooperative mechanisms, one being the effective action of the BCB population, which behaves as an electron reservoir continuously feeding the TSS bands \cite{Sobota-PRL-2012-Bulk-Reservoir, Gedik-2012-PRL-phonons}, the other one an electron bottleneck effect arising at the Dirac node, which is located at an energy of $E_{D}\sim$0.15 eV. Differently from previous findings \cite{Sobota-PRL-2012-Bulk-Reservoir, Reimann-2014-PRB-phonons, Hajlaoui-NatComm-2014-e-h-asym}, the bottleneck effect is surprisingly strong in our data, and manifests as a more pronounced time-dependent accumulation of intensity above the Dirac node, near the bottom of the upper Dirac cone in Fig. 1b. In contrast, the spectra further reveal a significant suppression of the Dirac-node spectral weight that is independent of $\Delta$t (see Supplementary). These effects can also be distinguished in measurements performed over a larger momentum region around the SBZ center (Figs. 1c and 1d). Here we show a full-ARPES mapping of the TSS measured at $\Delta$t=500 fs near the Dirac node (Fig. 1c), together with the corresponding constant-energy surfaces extracted in a wider energy range (Fig. 1d). The bottleneck effect emerges from a substantially reduced phase space for electron relaxation at the Dirac node. Such a blockade of the hot-electron decay is expected to be considerably strong in the presence of a band gap at the Dirac node, especially if the gap exceeds the maximum phonon energy available for electron relaxation ($\sim$25 meV). At first glance, this scenario might seem consistent with our data if we consider that, unlike linearly-polarized light, the rotating electric field of the circularly-polarized pump pulse might create a nonequilibrium massive-fermion state through the transient breaking of time-reversal symmetry \cite{Gedik-2013-Science-Floquet}. However, for this condition to be fulfilled, it is established that a strong coherent coupling between the intense laser field and the TSS is primarily required, implying that the pump-pulse photon energy must not exceed the size of the bulk band gap ($\sim$0.25--0.3 eV) \cite{Gedik-2013-Science-Floquet}. This limitation explains the absence of a dynamical gap at the Dirac node in Fig. 1b, a conclusion which is further supported by the absence of a bright Dirac node in recent tr-ARPES experiments using linearly-polarized 1.5 eV photons \cite{Sobota-PRL-2012-Bulk-Reservoir}. These findings motivate us to attribute the Dirac-node suppression in our data to final-state effects caused by matrix-elements introduced by the probe pulse. 

In this context, rather than coherent interaction with the TSS, our data reveal signatures of strong coupling between the circularly-polarized infrared pump pulse and bulk valence-band states located at an energy of  $\sim-$1 eV below the Fermi level (see Fig. 1e). This effect manifests itself as a pronounced intensity cut-off which remains at a constant energy of $\sim$0.5 eV above the Fermi level up to $\Delta$t=400 fs, before electrons from higher levels relax down below this energy. The energy difference (vertical red-solid arrow in Fig. 1e), exactly corresponds to the pump photon energy, indicating that both the TSS and BCB are largely populated through direct optical transitions from these deeper-lying levels (black-dashed line in Fig. 1e). Besides the Fermi level widening, other transitions at off-normal wave vectors contribute with a small weight only (vertical red-dashed arrow in Fig. 1e). Such direct interband excitation generates a high concentration of valence-band holes which rapidly relax within the subsequent dynamics. This process, which leads to an excess of hot electrons above the Fermi level, might represent another important contribution to the recently discovered electron-hole asymmetry in the Dirac cone \cite{Hajlaoui-NatComm-2014-e-h-asym}. 

Further insight into the dynamics of excited states can be obtained by analyzing the integrated tr-ARPES intensity within small energy-momentum windows (labelled from 1 to 17 in Fig. 1e). This method allows us to disentangle the TSS and BCB dynamics from the tr-ARPES spectra, as shown in Figs. 1f and 1g, respectively. As expected, the hot-electron avalanche within the TSS bands gives rise to a delayed electron filling and progressively slower electron decay when the energy decreases (Fig. 1f). A similar but less pronounced behavior can be observed for the BCB population (Fig. 1g). Once the BCB population reaches its maximum, both the TSS and BCB dynamics are very similar, resulting in decay times that follow each other upon reversal of the circular light polarization (Fig. 1h). This effect pinpoints bulk-to-surface coupling as one of the driving forces underlying the ultrafast electron relaxation. The strong coupling results in a slower electron thermalization and an average hot-electron temperature of $\sim$2.3$\times10^{3}$ K (Fig. 1i). The rise times within the higher-energy windows are longer than the time resolution, indicating that multiple elastic events of interband electron-electron scattering additionally play a role, in agreement with previous studies \cite{Sobota-PRL-2012-Bulk-Reservoir, Gedik-2012-PRL-phonons, Crepaldi-2012-PRB-phonons, Gedik-2013-Science-Floquet, Sobota-PRL-2013, Reimann-2014-PRB-phonons, Sobota-PRL-2014-Oscillations, Hajlaoui-NatComm-2014-e-h-asym}.  

Up to $\Delta$t$\sim$600 fs, the electron decay is dominated by intraband scattering, which leads to a momentum relaxation of the TSS and BCB populations from higher to lower energy, towards the bottom of each individual band. The subsequent dynamics mainly depends on the effective electron transfer from the BCB to the TSS through interband scattering. At $\Delta$t=950 fs, due to such a BCB-emptying process, a large portion of the BCB electrons has already decayed into the TSS bands (see Fig. 1b). Both interband and intraband scattering processes are strongly mediated by the emission of acoustic and optical phonons \cite{Optical-Orientation-1984, Othonos-JAP-1998, Sobota-JESRP-2014}. 

\begin{figure*}
\centering
\includegraphics [width=0.92\textwidth]{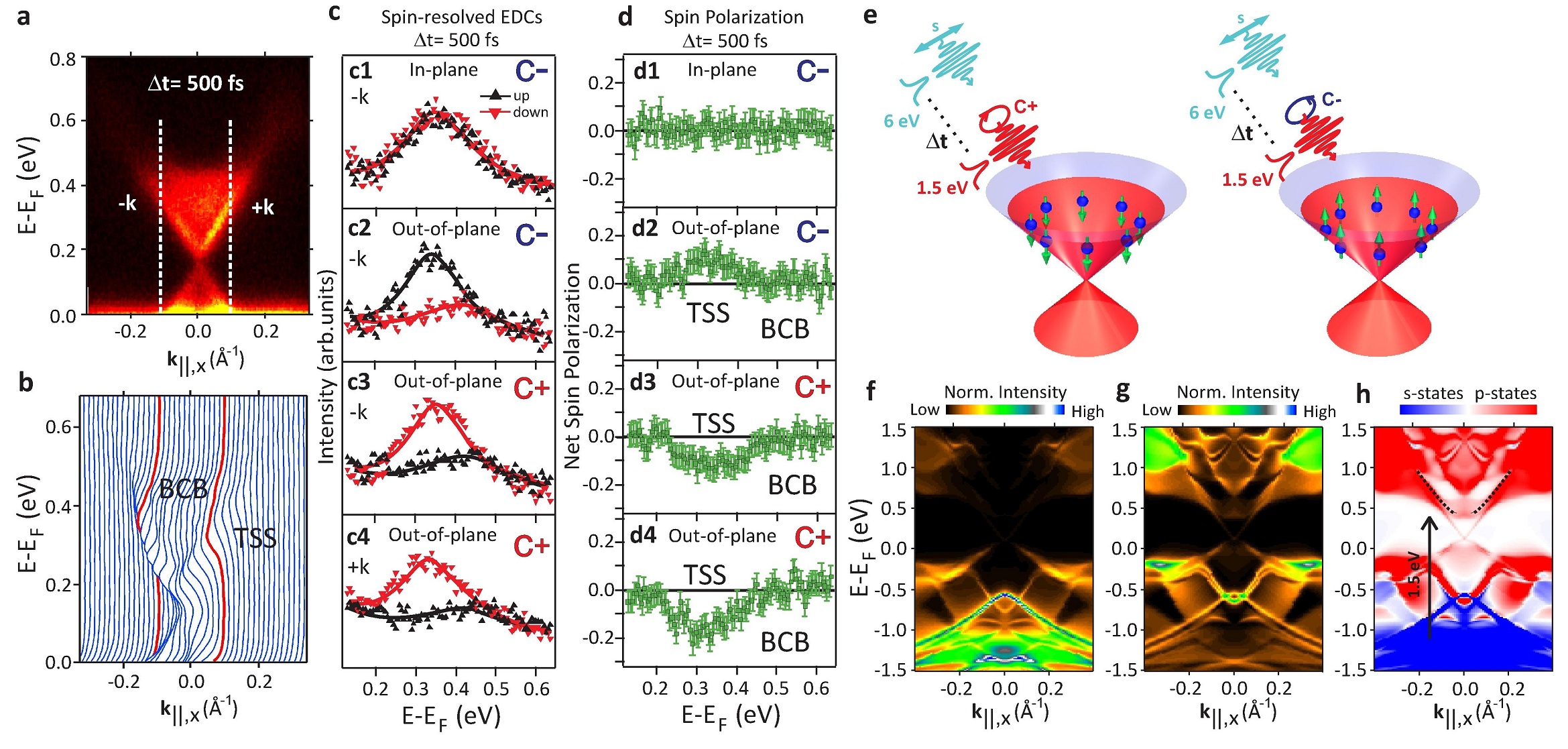}
\caption{{\bf Ultrafast spin-polarization control of Dirac fermions.} Spin-resolved tr-ARPES spectra are obtained along the \Gbar\Kbar\ direction upon reversal of the pump-pulse helicity from positive (C +) to negative (C-), at a fixed time delay of 500 fs. (a) Corresponding tr-ARPES dispersion indicating the momentum cuts at which spin-resolved measurements are presented (vertical-dashed lines). (b) Energy-distribution curves (EDCs) extracted from (a). EDCs corresponding to spin-resolved measurements are emphasized in red. (c) Spin-resolved spectra of the in-plane (c1) and out-of-plane (c2-c4) spin-polarization components obtained at opposite momenta using different light polarizations (C+ and C-). The out-of-plane projection is perpendicular to the surface normal, and the in-plane one tangential to the circular constant-energy contour of the TSS. (d1-d4) Corresponding net spin polarizations. The in-plane components associated to panels (c3, c4) and (d3, d4) provide similar results as in panels (c1) and (d1). (e) Simplified schematics of the ultrafast optical orientation process. (f-h) One-step photoemission calculations revealing dominant transitions from s-like initial states to p-like final states. In (f) and (g), p- and s-like initial-state orbitals are suppressed in the transition matrix elements, respectively. (h) The normalized intensity difference reveals distinct contributions from p- (red) and s-like orbitals (blue).}
\label{spin-control}
\end{figure*}

Remarkably, the transient TSS population in Fig. 1b is characterized by an asymmetric intensity distribution at opposite $+k_{\parallel,x}$ and $-k_{\parallel,x}$ wave vectors. The asymmetry originates from a circular-dichroic effect induced by the pump pulse polarization \cite{Mauchain-PRL-2013}. The dynamic recombination of excited BCB electrons within the Dirac cone leads to a progressive decay of this asymmetry with increasing time (see Supplementary), an effect that is the manifestation of a time-dependent electrical current or, in other words, a transient net flow of charge across the sample surface. The overall decay of the generated current is slightly less than the relaxation time of the total population, and we will return to this point later. At the onset of the optical excitation, such an imbalance in the density of excited carriers is equivalent to a circular-dichroic effect in the Dirac bands that reaches an absolute value of $\sim$15\% upon reversal of the pump-pulse polarization from C+ to C- (see Supplementary for more details). This effect, not present in the BCB, indicates that reversal of the pump-pulse helicity reverses the direction of the electric current, and that this phenomenon occurs only on the surface. These findings clearly demonstrate previous conclusions from transport experiments \cite{Gedik-2012-NatNanotech-photocurrents} and differ from tr-ARPES studies using 1.5 eV photons, so far conducted with linearly-polarized pump pulses \cite{Sobota-PRL-2012-Bulk-Reservoir, Gedik-2012-PRL-phonons, Crepaldi-2012-PRB-phonons, Sobota-PRL-2013, Sobota-PRL-2014-Oscillations, Hajlaoui-NatComm-2014-e-h-asym}. In this respect, we point out that a fully-symmetric transient TSS population with respect to the surface normal, such as the one generated using linearly-polarized infrared light \cite{Sobota-PRL-2012-Bulk-Reservoir}, would not give rise to the ultrafast electrical current observed here. The reason is that an entire suppression of the net flow of charge on the surface occurs if, for all values of $\Delta$t, there is an equal number of excited electrons moving in opposite momentum directions. 

The overall relaxation dynamics observed here proceeds in a much faster timescale than the one previously found in other prototypical TIs such as \BiSe\ \cite{Sobota-PRL-2012-Bulk-Reservoir} and \BiTe\ \cite{Hajlaoui-NatComm-2014-e-h-asym}, where the nonequilibrium TSS population exhibited long lifetimes of more than 10 and 50 ps, respectively. We attribute this significant difference to the fact that, in the case of \SbTe, we observe a smaller separation in energy-momentum space between the TSS and BCB populations. This situation facilitates to a large extent phonon-mediated transfer of BCB electrons into the TSS bands, leading to a stronger bulk-to-surface coupling and thus to a pronounced bottleneck effect through an accelerated electron dynamics.\\

\noindent{\bf All-optical sub-ps control of the spin polarization}

\noindent{To} investigate the transient response of the Dirac-cone spin polarization to circularly-polarized infrared light, we perform spin-resolved tr-ARPES measurements at a fixed time delay of 500 fs upon reversal of the pump-pulse helicity from C+ to C- (Fig. 2). The tr-ARPES dispersion corresponding to this time delay is shown in Fig. 2a, where vertical-dashed lines indicate the electron wave vectors at which spin-resolved measurements are presented. The spin-integrated energy-distribution curves (EDCs) extracted at these wave vectors are emphasized in red in Fig. 2b. Each EDC contains a pronounced TSS peak located at an energy of $\sim$0.33 eV, where the Dirac-cone distortion due to hexagonal warping is rather weak (see Figs. 1c-1d). This peak is accompanied by a less intense but clearly distinguishable high-energy tail due to the BCB contribution. 

Our spin-resolved measurements (Figs. 2c and 2d) reveal that circularly-polarized infrared photons transiently flip the photoexcited TSS spin polarization perpendicular to the surface, and that reversing the pump-pulse helicity reverses the direction of the resulting out-of-plane spin polarization. These effects are clearly observed for a fixed $k_{\parallel,x}$ wave vector in the in-plane and out-of-plane spin-resolved EDCs shown in Figs. 2c1-2c3, and in the corresponding net spin polarizations (Figs. 2d1-2d3). Specifically, the in-plane component of the TSS spin polarization is nearly zero (Fig. 2d1), and the out-of-plane one reaches a maximum absolute value of $\sim$12\% upon reversal of the pump-pulse helicity from C- to C+ (Figs. 2d2 and 2d3). Moreover, the out-of-plane spin polarization does not reverse when going from $-k_{\parallel,x}$ to $+k_{\parallel,x}$ wave vectors (Figs. 2c4 and 2d4). These findings establish that the ultrafast net flow of charge on the surface is accompanied by a fully controllable spin polarization. 

In contrast, the transient BCB population contributes negligibly to the measured spin polarizations. This result is consistent with previous conclusions from time-resolved reflectivity experiments using infrared light \cite{Gedik-2011-PRL-Kerr}, and with the fact that due to bulk-inversion symmetry, the BCB population is expected to be essentially unpolarized. The effect is seen in our data as a nearly-zero spin polarization of the high-energy BCB tail above $\sim$0.4 eV. Although the pump-pulse helicity might initially create spin-polarized BCB carriers, it has been shown that due to strong spin-orbit coupling, their spin polarization will rapidly decay to zero faster than the time resolution of the present experiment \cite{Gedik-2011-PRL-Kerr}. On the other hand, the impact of the spin-orbit interaction on the TSS is expected to cause spin relaxation rather than decoherence of its transient spin polarization. In particular, the theoretically expected spin-relaxation time of the TSS in prototypical TIs can exceed several ps \cite{Liu-PRL-2013, Zhang-PRB-2013}. The fact that the BCB in Fig. 2 is already unpolarized after 500 fs indicates that in the presence of strong spin-orbit coupling, electron-electron scattering acts as the dominant depolarizing mechanism for bulk states. This finding implies that the electron-electron scattering rates for bulk and surface states are different by at least one order of magnitude. We note that this difference is not necessarily connected to an elevated hot-electron temperature of the BCB population. Moreover, differently from the TSS, the BCB spin polarization is prone to be affected by electron-impurity scattering events following optical excitation. Other effects associated with spatial diffusion of BCB electrons are likely to play a minor role \cite{Sobota-PRL-2012-Bulk-Reservoir}.

It is difficult to directly compare the observed magnitude of the transient TSS spin polarization in Fig. 2 to the one expected from ground-state model calculations \cite{Pauly-PRB-2012, Yazyev-theory-PRL-2010}, as the spin polarization in the nonequilibrium state can be strongly affected by multiple-scattering events that occur during electron thermalization and within the subsequent hot-electron dynamics, resulting in a substantial spin depolarization \cite{Gedik-2011-PRL-Kerr}. For instance, the TSS spin polarization can be reduced within the timescale of electron-electron scattering and below the time resolution of the present experiment depending on the rate of electron-electron collisions \cite{Gedik-Nature-2DEG-2005}, and without affecting the transient charge current on the surface. Elastic events of interband electron-electron scattering giving rise to an enhanced bulk-surface coupling upon electron thermalization may additionally act as spin-depolarization channels \cite{Gedik-2011-PRL-Kerr}. However, in Fig. 2, we find a non-zero value of the transient TSS spin polarization, indicating that the fastest depolarization events associated with elastic electron-electron scattering are not sufficient to completely suppress the spin polarization. The same is true for ultrafast spin-depolarization processes associated with an hyperfine interaction between excited electrons and nuclear spins \cite{Optical-Orientation-1984}, indicating that the hyperfine coupling is too weak to transiently create an internal magnetic field.
\begin{figure*}
\centering
\includegraphics [width=0.9\textwidth]{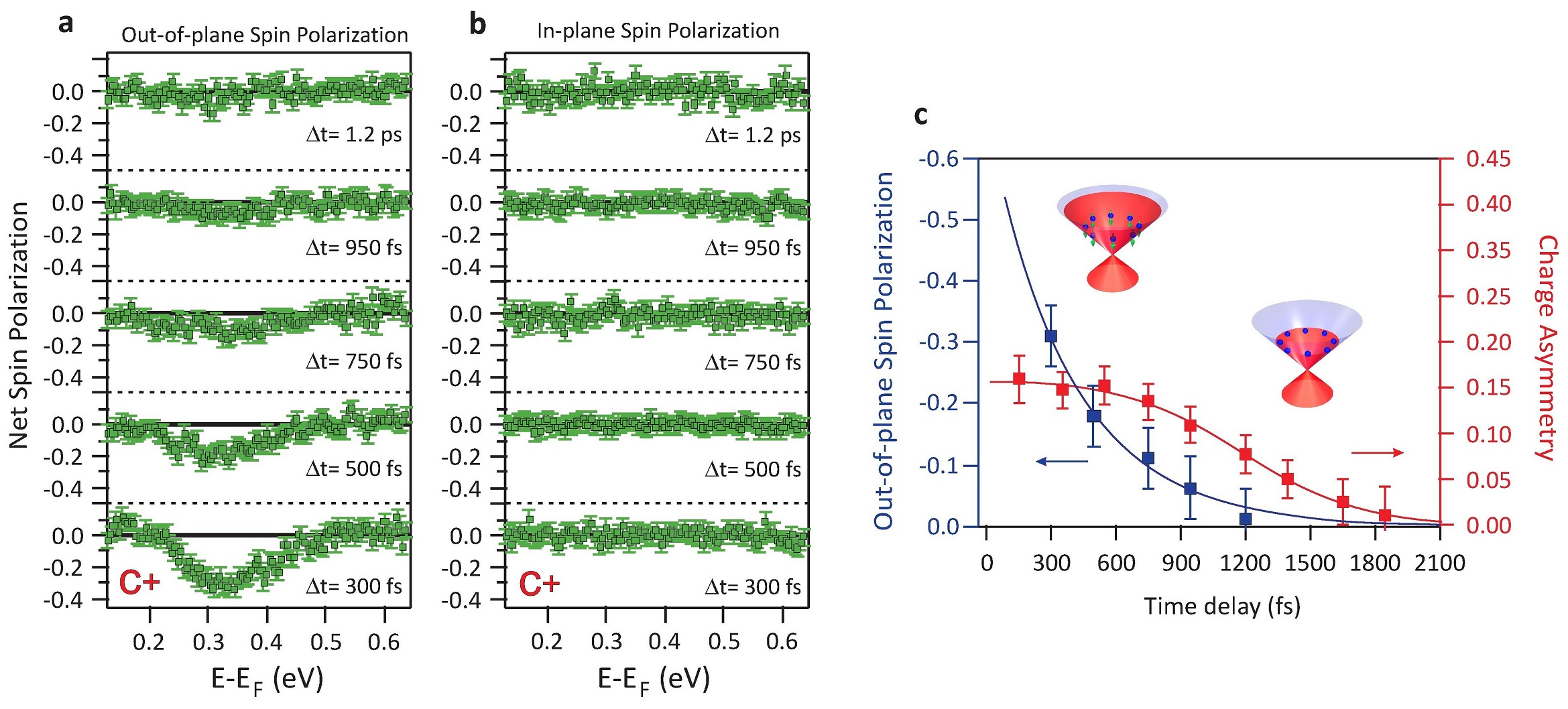}
 \caption{{\bf Probing the spin-polarization and charge dynamics of Dirac fermions.} (a,b) Ultrafast temporal evolution of the TSS (a) out-of-plane and (b) in-plane components of the spin polarization following photoexcitation with circularly-polarized infrared pulses of positive helicity (C+). Similar to Fig. 2, the out-of-plane spin projection is perpendicular to the surface normal, and the in-plane one tangential to the circular constant-energy contour of the TSS. (c) The relaxation dynamics of the net out-of-plane spin polarization (blue) and charge asymmetry (red) exhibits different timescales. The spin polarization relaxes with a time constant of about 0.4 ps, while the charge asymmetry is delayed by $\sim$1 ps.}
\label{spin-dynamics}
\end{figure*}

Another depolarization pathway is intraband decay of unpolarized electrons from energies higher than $\sim$0.6 eV, where the TSS and BCB connect to each other bridging the gap of the volume \cite{Seibel-PRL-2015}. This depolatization pathway has been proved very effective in recent spin-resolved tr-ARPES experiments focusing on the surface resonances of \BiSe\ using linearly-polarized light \cite{Cacho-Surface-resonances-PRL-2014}. However, this contribution seems to be small in our data, as most of the excited electrons are generated below the intensity cut-off observed at $\sim$0.5 eV in Figs. 1b-1e. Moreover, spin depolarization of the Dirac cone can also occur upon momentum relaxation at every stage of the hot-electron avalanche due to scattering with phonons \cite{Optical-Orientation-1984}.

In Fig. 2e, we summarize the transient behavior of the Dirac-cone spin polarization by showing a simplified scheme of the optical spin orientation process. The circular pump pulse aligns the hot-electron spin polarization antiparallel to the surface normal for C+ and along the opposite direction for C- light, leading to a transient spin texture that is similar to the one recently predicted \cite{Louie-PRL-2012-Theo-Spin-Flip} and observed using circularly-polarized 6 eV photons in equilibrium \cite{Jozwiak-NP-2013-Spin-flip, Sanchez-Barriga-PRX-2014}. The difference in the present case is that the spin-reorientation process occurs for nonequilibrium conditions and is solely caused by the pump-pulse helicity in sub-ps timescales. We emphasize that after the first photoemission step induced by the pump pulse, the linear polarization of the 6 eV probe pulse is perpendicular to the out-of-plane TSS spin polarization in the intermediate state, thus leaving the photoelectron spin polarization unchanged upon the second excitation above the vacuum level (see Supplementary for details). The spin-reorientation process induced by the pump pulse can be viewed in analogy to the circular-photogalvanic effect previously observed in semiconductor quantum wells \cite{Ganichev-Nature-2002}. The process is the result of optical spin pumping into the unoccupied Dirac cone with electrons that are excited with preferential spin orientations. The resulting TSS out-of-plane spin polarization ultimately originates from the spin dependence of the relativistic dipole matrix elements in the presence of strong spin-orbit coupling, which in turn depend on the symmetry of the states involved in the optical transition \cite{Sanchez-Barriga-PRX-2014}. 

Hence, to examine the wave function character of the states that contribute to the spin-reorientation process, we perform one-step model photemission calculations under equilibrium conditions using linearly s-polarized 6 eV photons (Figs. 2f-2h). We analyze the symmetry of the wave functions by suppressing p- or s-like initial-state orbitals in the corresponding transition matrix elements and normalize the intensity (Figs. 2f and 2g, respectively). While the TSS contains contributions from all p-orbitals as expected \cite{Pauly-PRB-2012, Yazyev-theory-PRL-2010}, the calculations reveal that the transient TSS spin polarization originates predominantly through direct optical transitions from spin-degenerate bulk states of s-like character (Fig. 2h). These states are thus represented by an initial-state wave function of high symmetry (angular momentum $l$=0), which is a precondition for optical spin orientation in solid surfaces using circularly-polarized light \cite{Sanchez-Barriga-PRX-2014}. The calculations also reveal bulk-like surface resonances located at the BCB edges (black-dashed lines in Fig. 2h). We emphasize that the wave function of the TSS consists of a complex superposition of different spin directions appearing on different atoms \cite{Pauly-PRB-2012, Yazyev-theory-PRL-2010}. As a consequence, the out-of-plane TSS spin polarization shown in Figs. 2d3 and 2d4 reveals only weak momentum dependence at opposite wave vectors, and preserves time-reversal symmetry. This process can be understood in terms of orbital angular-momentum transfer from the circularly-polarized pump pulse, which in the presence of strong spin-orbit coupling triggers spin-flip transitions as a function of the light helicity.  Moreover, the magnitude and direction of the spin polarization are expected to be time-dependent \cite{Liu-PRL-2013, Zhang-PRB-2013}, implying in this way that transient changes in the relative orbital occupations are equally possible.\\ 
\begin{figure*}
\centering 
\includegraphics [width=0.85\textwidth]{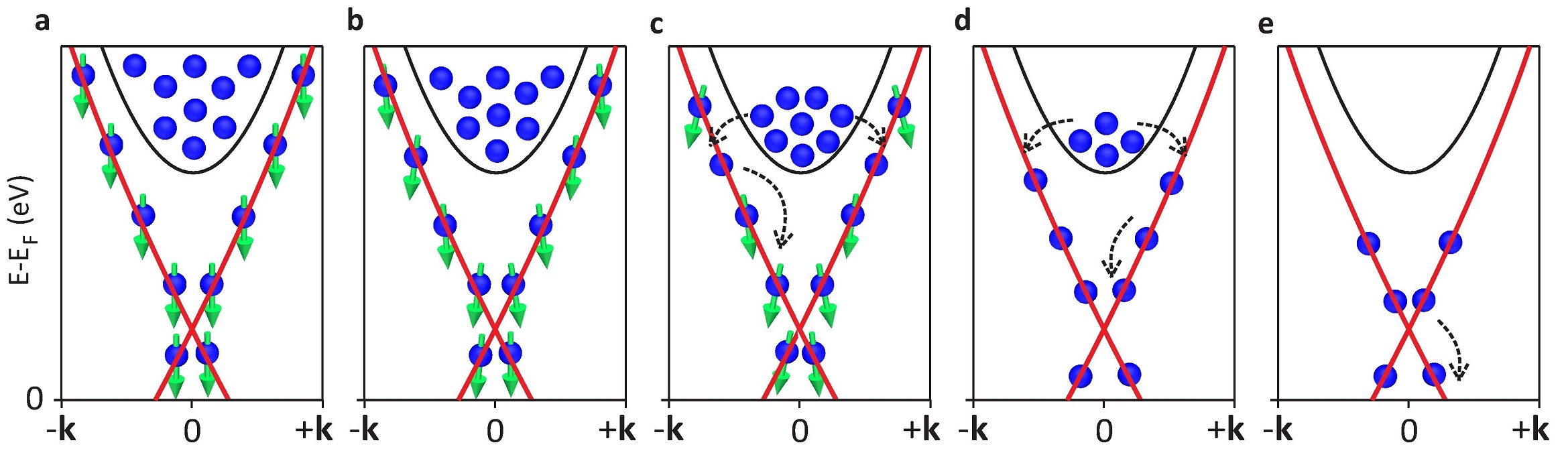}
\caption{{\bf Schematics of the microscopic mechanisms underlying the ultrafast relaxation of the Dirac-cone spin polarization.} From (a) to (d), subsequent stages of the spin-relaxation process are shown. The straight red lines and the parabolic black curve represent the TSS and BCB above the Fermi level, respectively. Blue filled circles denote photoexcited electrons, and green arrows correspond to the out-of-plane spin polarization. (a) A circular pump-pulse of positive helicity (C+) creates a thermalized spin-polarized population of hot electrons in the TSS. (b) The initial relaxation of the TSS spin polarization mainly proceeds through intraband scattering processes. (c) Once electrons within the BCB decay to their band minimum, the unpolarized BCB electron population is transferred into the Dirac bands via phonon-assisted decay, so that bulk-to-surface coupling further contributes to the spin depolarization. (d, e) Subsequently, this process continues until the Dirac cone is largely populated by BCB electrons, leading to a suppression of the charge current on the surface (see text).}
\label{Spin-relaxation-process}
\end{figure*}

\noindent{\bf Ultrafast spin-polarization dynamics}

\noindent{To} understand the spin-polarization dynamics of the TSS, in Fig. 3 we monitor the temporal evolution of the out-of-plane (Fig. 3a) and in-plane (Fig. 3b) components of the spin polarization after optical excitation with circularly-polarized infrared pulses of positive helicity (C+). Similar to Fig. 2, the measurements are taken at a fixed $+k_{\parallel,x}$ wave vector and along the \Gbar\Kbar\ direction of the SBZ. We observe a rapid and progressive suppression of the out-of-plane spin polarization, while the in-plane component remains nearly zero. After $\sim$~1.2 ps, the out-of-plane spin polarization is strongly reduced, implying that spin decoherence due to scattering with phonons rather than a gradual recovery of the expected Dirac-cone helical spin texture likely dominates the momentum-relaxation process. Moreover, the fact that the spin polarization disappears faster than the TSS population reveals that the transient spin-polarized electrical current originating from Dirac fermions transforms into a pure electrical one. 

In Fig. 3c, we gain further insight into this transformation by comparing the  out-of-plane spin-polarization dynamics with the one of the transient charge asymmetry, which is extracted at selected time delays from the normalized intensity imbalance between the TSS bands at opposite $+k_{\parallel,x}$ and $-k_{\parallel,x}$ wave vectors (see Supplementary). The asymmetry is evaluated at an energy of $\sim$0.3 eV, in correspondence with the spin-resolved measurements of Figs. 3a and 3b. We clearly observe that the timescales associated to spin decoherence and relaxation of the charge asymmetry within the Dirac cone are drastically different. In particular, up to $\sim$600 fs, while the spin polarization is considerably reduced, the charge asymmetry remains nearly constant. This behavior reveals that intraband scattering is a spin-depolarizing process that develops under the preservation of the transient charge current on the surface. At the same time, the electron bottleneck effect arising at the Dirac node slows down the depolarization process while maintaining the charge asymmetry unchanged. After $\sim$600 fs, once the BCB electrons start to more effectively decay into the TSS bands trough interband scattering, the spin polarization is further reduced, and the charge asymmetry continuously decreases until it nearly disappears within $\sim$1.85 ps. Note that the decay time is slightly shorter than the $\sim$2.5 ps needed for the TSS population to relax below $\sim$0.3 eV (see Fig. 1f), suggesting that the quenching of the charge current occurs once the Dirac cone is largely filled by electrons originating from the BCB. From these results we derive that the slow charge-current decay is directly caused by the interplay between the electron bottleneck arising at the Dirac node and BCB-emptying processes, which simultaneously act as spin-depolarization channels. Hence, the suppression of the spin polarization is caused by the cooperative action of intraband and interband scattering processes. We emphasize that in a nonequilibrium massive-fermion state, we would have expected the charge current to immediately drop within the electron-electron scattering time due to elastic backscattering under 180 degrees, differently from what is observed here. The slow charge dynamics in Fig. 3c additionally suggests that a transient gap at the Dirac node does not open during consecutive momentum-relaxation stages. This fact is consistent with our general interpretation concerning the spectral weight suppression at the Dirac node.

Based on this conclusion, in Fig. 4 we illustrate a simplified scheme that focuses only on the mechanisms underlying the spin-relaxation process. Up to $\sim$600 fs, the oriented spin-polarized TSS population (Fig. 4a) decays mainly through intraband scattering (Fig. 4b). The electron momentum relaxation in the presence of spin-orbit coupling results in small random rotations of the electron spins due to scattering with phonons. In this process, the spin-orbit field can be viewed as an effective magnetic field acting on the electron spins. Every scattering event results in a randomization of the electron spin-precession axis around this field, causing spin depolarization. Subsequently, after the BCB electrons have reached their band minimum, the TSS spin polarization additionally decays through interband scattering (Fig. 4c). We emphasize that because the TSS and BCB edges are close to each other in energy-momentum space, interband scattering is likely to occur before the BCB electrons decay to their band minimum, but in a less effective way. The process can be viewed as transfer of unpolarized electrons into the TSS bands, implying that bulk-to-surface coupling and intraband scattering cooperatively depolarize the electron spectrum within $\sim$1.2 ps (Fig. 4d). The subsequent dynamics proceeds once the TSS population is completely filled with unpolarized BCB electrons at $\sim$1.85 ps (Fig. 4e). These electrons remain unpolarized throughout the decay towards the Fermi level due to the strong contribution of unpolarized bulk-valence band states located right below and in the immediate vicinity of the Fermi level (see Fig. 1e). 

Our results establish that the transfer of BCB electrons into the TSS is a spin-depolarizing event, an effect that we attribute to the complex nature of the phonon-mediated interaction between the bulk and surface electronic wave functions. The partial filling of the TSS bands with unpolarized electrons respects the total wave function symmetry of the TSS, as it is composed by different spin orientations on different atoms \cite{Pauly-PRB-2012, Yazyev-theory-PRL-2010}. Our photoemission calculations in equilibrium additionally reveal that the strength of the bulk-surface interaction is possibly mediated by the surface resonances appearing at the BCB edges \cite{Seibel-PRL-2015}. These observations recall a strong need for refined spectroscopical calculations of time-resolved photoemission spectra, which quantitatively describe dynamical processes as a function of time.  A first  theoretical description of time-resolved photoemission for real systems was introduced recently \cite{Braun-PRB-2015}. Although the numerical implementation will need some time, at present such type of calculations are not possible using only perturbative methods. It would be interesting to perform similar studies as the one presented here using lower pump photon energies so that all electrons are excited within the bulk band gap, or to use more bulk-insulating samples. Another alternative is to use ultrathin TI films without the contribution from the bulk conduction. 

To summarize, we have shown that circularly-polarized infrared pulses of opposite helicities allow all-optical control of the nonequilibrium spin polarization of TSSs on ultrafast timescales. The dynamics of the generated spin-polarized electrical currents exhibits two distinct timescales for the spin and charge degrees of freedom on the surface. We have provided key evidence for the mechanisms underlying this behavior. Our observations establish the utilization of photoexcited TSSs as unique channels in which to drive ultrafast currents with fully controllable spin configurations. The results presented here are of critical importance in understanding the spin properties of TIs following a perturbation by intense laser fields and open the way for novel applications of these materials in ultrafast spintronics.\\

\noindent {\bf Methods}

\noindent{Spin-resolved} tr-ARPES experiments are performed at room temperature in ultrahigh vacuum better than $1\cdot10^{-10}$ mbar. We use a Mott-type spin polarimeter coupled to a Scienta R8000 hemispherical analyzer, allowing us to detect the out-of-plane and one tangential component of the spin polarization (see Supplementary for more details). As pump and probe pulses we use the first and fourth harmonics derived from a Coherent RegA Ti:Sapphire amplified laser operating at 150 kHz. \SbTe\ single crystals are grown by the Bridgman method and cleaved {\it in situ}. Photoemission calculations are based on multiple scattering theory within the one-step model of photoemission including wave-vector, spin and energy-dependent transition matrix elements \cite{Hopkinson-CPC-1980, Braun-theory-96}. We use a fully-relativistic version that is part of the SPR-KKR program package \cite{Ebert-SPRKKR-2011,Ebert-SPRKKR-2012}, with spin-orbit coupling included self-consistently.  


\noindent {\bf Acknowledgements}. Financial support from the Deutsche Forschungsgemeinschaft (Grant No. SPP 1666) and the Impuls-und Vernetzungsfonds
der Helmholtz-Gemeinschaft (Grant No. HRJRG-408) is gratefully acknowledged. J. M. is supported by the CENTEM project CZ.1.05/2.1.00/03.0088, co-funded by
the ERDF as part of the Ministry of Education, Youth and Sports OP RDI programme.

\noindent {\bf Author contributions}. J. S.-B and E. G performed the experiments with assistance from A. V., O. K. and R. S.; L. Y. provided bulk-single crystals and performed sample characterization; J. B., J. M., H. E. carried out calculations; J. S.-B, E. G and O. R. performed data analysis, figure and draft planning; J. S.-B. wrote the manuscript with input from all authors; J. S.-B. and O. R. were responsible for the conception and the overall direction.

\noindent {\bf Additional information}. Supplementary Information accompanies this paper is available at http://****. The authors declare no competing financial interests. Correspondence and requests for materials should be addressed to J. S.-B. (Email: jaime.sanchez-barriga@helmholtz-berlin.de).














\end{document}